\documentstyle[preprint,prd,tighten,eqsecnum,aps,epsf]{revtex}
\begin{document}
\preprint{\vbox{
\hbox{SUNY-NTG-94-41}
\hbox{August 1994}
\hbox{Phys. Rev. D {\bf 50}, 6824 (1994)}
\hbox{      }\hbox{      }
\hbox{       }\hbox{      }
}}
\draft
\title{\bf
Consistency of data on soft photon production\\
in hadronic interactions\\}
\author{Peter Lichard\footnote{
On leave of absence from the
Department of Theoretical Physics, Faculty of Mathematics and Physics,
Comenius University, SK 842-15 Bratislava, Slovakia.
}}
\address{Department of Physics, State University of New York
at Stony Brook,\\
Stony Brook, New York 11794}

\maketitle
\begin{abstract}
The glob model of Lichard and Van Hove and the modified soft
annihilation model (MSAM) of Lichard and Thompson are used as
a phenomenological tool for relating results from various experiments
on soft photon production in high energy collisions. The total
phenomenological expectation is composed of contributions from
classical
bremsstrahlung, the soft annihilation model and the glob model.
The empirical excess above the background from hadronic decays at very
small longitudinal momenta of photons is well reproduced, as well as
that for transverse momenta $p_T\gtrsim 10$ MeV/$c$. Some data do not
require the glob model and MSAM components in the phenomenological
mixture, but do not exclude them. On the basis of consistency of all
data with the total theoretical
expectation we argue that the results of all experiments are mutually
consistent. The models are unable to describe the excess of ultrasoft
photons ($p_T \lesssim$ 10 MeV$/c$), seen by some, but not all,
experiments. This may indicate an as yet unknown 
projectile-mass-dependent
production mechanism. Possible relations of soft photon production to
other phenomena are discussed. A simple-to-use, but physically
equivalent version of the glob model is developed, which enables an
easy check of presented results.
\end{abstract}

%\pacs{PACS number(s): 13.85.Qk, 12.40.-y, 12.38.Bx }
% incl.prod.with ident.leptons,photons or other nonhad part's
% other models of strong interactions
% QCD-perturbative calculations

\narrowtext
\newpage
\section{INTRODUCTION}
\label{intro}

The inclusive production of soft photons in particle collisions at
high energies has been studied in several experiments
\cite{SLAC,BEBC,AFS,AFS1,EMC,HELIOS,NA22,SOPHIE,antos} using various
experimental techniques.

In a pioneering bubble chamber experiment performed at the Stanford
Linear Accelerator Center (SLAC), Goshaw {\it et al.} measured the
transverse momentum distribution of photons produced in $\pi^+p$
collisions at 10.5~GeV/$c$. Using their own data about hadronic final
states they showed that the bremsstrahlung formula of classical
electrodynamics \cite{jackson} was able to account for all the
observed photon yield.

A few years later, the experimental group led by Yves
Goldschmidt-Clermont \cite{BEBC} investigated the photon production
in 70~GeV/$c$ $K^+p$
collisions using the Big European Bubble Chamber (BEBC) at the CERN
Super Proton Synchrotron (SPS). In contrast to the observation of the
SLAC experiment they found many more photons with very small
longitudinal momentum ($2|p_L|/\sqrt{s}\lesssim 0.005$)
than they expected on the basis of the classical bremsstrahlung
formula and their charged hadron production data.  Similarly, the
unpublished \cite{BEBCpt} photon transverse momentum spectrum from
the same experiment exhibited a significant excess over the
bremsstrahlung estimate for $p_T\lesssim$ 60 MeV/$c$.

The results obtained by the Axial Field Spectrometer (AFS)
collaboration at the CERN Intersecting Storage Rings (ISR) \cite{AFS}
were in conformity with both previous experiments. They neither
required an additional source to explain the observed photon signal,
nor excluded
an anomalous component of the size indicated by the BEBC experiment.
But they did rule out a strong increase of anomalous photon production
with rising collision energy (the invariant energy $\sqrt s$ in the
AFS experiment was 63 GeV, compared to 11.5 GeV in the BEBC
experiment).

In a subsequent AFS experiment \cite{AFS1}, the direct photon
production was studied in $pp$ and $\alpha\alpha$ collisions
at $\sqrt s=63$~GeV. The photon momentum range explored
(0.1~GeV/$c<p_T< 1$~GeV/$c$) does not overlap the range in this paper
($p_T\lesssim 0.1$~GeV/$c$). This is the reason why we will not 
include these data in our comparison. Let us only note that
the results of \cite{AFS1} did not show any excess
beyond the limits dominated by systematic uncertainties.

The inclusive yield of photons from deep inelastic $\mu p$ scattering
at 200 GeV was measured by the European Muon Collaboration (EMC) at
the CERN SPS \cite{EMC}. After subtracting the contributions from
hadron electromagnetic decays and Bethe-Heitler muon bremsstrahlung,
residual photons were observed at a mean level of $0.15\pm 0.06$ per
event.

The first results of the HELIOS (NA34) experiment at the CERN SPS
\cite{HELIOS} on soft photon production in $p$-Be and $p$-Al
interactions at 450 GeV/$c$ indicated a marked excess in the region
4~MeV/$c < p_T <$ 20~MeV/$c$. In addition, the yield seemed to
increase approximately as the square of the associated hadron
multiplicity, which would have signified a collective production
mechanism \cite{square}. These findings were not confirmed by a more
extensive later experiment~\cite{antos} by some members of the same
group.

In the experiment performed by the EHS-NA22 Collaboration \cite{NA22}
at the CERN SPS,
the European Hybrid Spectrometer (EHS) was equipped with the Rapid
Cycling Bubble Chamber (RCBC) as vertex detector. The inclusive soft
photon production was studied in $\pi^+p$ and $K^+p$ collisions at
250 GeV/$c$. The results confirmed the existence of an anomalous
prompt photon signal with very similar properties as seen in the
BEBC experiment \cite{BEBC,BEBCpt}.

The SOPHIE/WA83 experiment \cite{SOPHIE}, which was performed at the
CERN SPS, used the OMEGA spectrometer supplemented with two
electromagnetic calorimeters. It differed in three important respects
from the previous two hadronic collision experiments
\cite{BEBC,NA22} that saw the anomalous soft photon signal.
First, it used a beam of negative particles
(pions with momentum of 280 GeV/$c$). Second, it was a purely
electronic experiment without utilizing a bubble chamber. Third,
it explored a kinematic region where the contribution of gammas from
hadronic decays is relatively small. The fact that the results of this
experiment, as concerns the existence and approximate magnitude of the
anomaly, are in conformity with those of previous ones
\cite{BEBC,NA22} is therefore especially valuable.

Anto\v{s} {\it et al.} \cite{antos} used a modified setup of the
HELIOS experiment \cite{HELIOS} and measured the inclusive $p_T$
spectra of soft photons produced at central and slightly backward
rapidities in 450 GeV/$c$ $p$-Be collisions.
Two independent photon measurement methods with corresponding
detectors and analysis chains were used in parallel: (i) a combination
of gas chambers, converter plates and a BGO matrix for the conversion
method; (ii) a BaF$_2$ array for the time of flight photon
identification method. The authors of \cite{antos} observed a
significant excess of direct photons at very low $p_T$ ($<15$~MeV/$c$)
above the background from hadronic decays. This excess was consistent
with the expected contribution from hadronic bremsstrahlung,
calculated from the classical electrodynamics formula \cite{jackson}.

On the theoretical side, the results of the EMC photon experiment
\cite{EMC} were described \cite{EMCSAM} using the soft annihilation
model \cite{SAM,SAM81}. Some discrepancy between experimental and
theoretical spectra stimulated the creation of the modified soft
annihilation model \cite{MSAM}, which allowed, contrary to
the original model, also gluons in the intermediate parton state.
We will present this model in some detail in
Sec.~\ref{MSAM}. Every attempt to explain the photon excess seen
in the BEBC experiment \cite{BEBC} on the same footing remained
unsuccessful.

To our knowledge, the first theoretical article that addressed the
issue of very soft anomalous photons \cite{BEBC} was that by
Van~Hove \cite{vanhove}.
He offered a common explanation of several ultrasoft phenomena
observed experimentally in central rapidity regions of high energy
collisions. These phenomena are characterized by very low
transverse momenta (production of soft pions or photons) or very
short distances in rapidity (fluctuations in rapidity
distribution--``intermittency"). Van Hove argued that they could be
regarded as manifestations of the occurrence, in at least some of the
collisions, of an intermediate parton system with considerable
lifetime and spatial extension. This system carries only a part of
the collision energy and momentum and the rest of the event is
``standard". The most natural framework for understanding the
production mechanism and properties of such a system is provided by
the QCD parton shower model \cite{shower}. It is usually assumed that
the shower development stops when parton virtualities fall to
$Q_0\approx 1$ GeV. The partons with this virtuality are supposed to
enter the hadronization process. Van Hove proposed that some of them
may continue in showering, producing a large multiplicity
($N_p\approx 30$) system of very soft partons, a glob of cold
quark-gluon plasma. The intermediate parton system consists of one or
several globs with masses $M_G\approx 1$ GeV/$c^2$. As discussed
in~\cite{vanhove}, globs need much time to hadronize, because
hadronization requires a very drastic rearrangement of partonic
wave functions. During their long lifetime, globs can produce photons
in the subprocesses $q + \overline{q}\rightarrow \gamma + g$ and
$g + q(\overline q) \rightarrow \gamma + q (\overline q)$. A
quantitative model for soft photon production based on the Van~Hove
ideas \cite{vanhove} was constructed and successfully compared
to the BEBC data \cite{BEBC,BEBCpt} by him and the present author
in~\cite{glob}.

Barshay \cite{barshay} investigated the role of pion condensation
\cite{migdal,brown} in soft photon production. He showed that this
collective coherent mechanism implied that the photon emission was
proportional to the square of the associated pion multiplicity. At
that time, this feature seemed to be indicated by the preliminary
HELIOS data \cite{HELIOS}.

Shuryak \cite{shuryakrefl} showed that the backward reflection of
pions at the boundary of a hadronic system, induced by the
modification of the pion dispersion curve by many-particle
interactions, strongly increases soft photon emission. A complex
view of the dense interacting pion matter was presented in the
subsequent work \cite{shuryakdrops} together with the implications
for experimentally observed phenomena. The photon yield was calculated
by applying the classical bremsstrahlung formula along the paths
of many times rescattered pions. As stated in the original
paper~\cite{shuryakdrops}, this approach could not describe the
observed excess of real photons at low transverse momenta.

V.~Balek, N.~Pi\v{s}\'{u}tov\'{a}, and J.~Pi\v{s}\'{u}t
\cite{balekapp} (a more detailed discussion of some issues can be
found in their later papers: \cite{balek1} in collaboration with
Zinovjev, and \cite{balek2}), first briefly reviewed experimental
information on, and theoretical understanding of the production of
very soft photons. Then they studied some effects which were
not considered in the bremsstrahlung calculation of the HELIOS
Collaboration \cite{HELIOS} and found them very small.
They also suggested two photon production mechanisms:
1) shock waves in the system of final state hadrons,
2) bremsstrahlung emitted by a quark that tries to
escape from the intermediate parton system.
They concluded that none of them could explain the
soft photon anomaly observed in experiments \cite{BEBC,HELIOS}.

Czy\.{z} and Florkowski \cite{czyz} used the classical
bremsstrahlung formula to calculate soft photon emission within the
framework of the boost-invariant color-flux tube model \cite{tube}.
The ratio of their photon production rate to the classical
one depends strongly on the direction of the photon and is very
sensitive to the assumed mass of quarks. For photons emitted
perpendicular to the collision axis and $m_q \approx10$ MeV/$c^2$
it can reach 10. The strong angular dependence is a feature that
saliently distinguishes this model from the essentially isotropic
glob model \cite{glob}.

Only a few of the theoretical approaches mentioned above
have attempted to make a detailed comparison with actual experimental
data, including the very important and
delicate matter of experimental acceptance and cuts.

While technical and experimental problems are substantial, an
important problem in assessing the reality of the anomalous photon
production is that the experiments have been performed under very
different conditions. The varied experimental conditions have included
different projectiles and targets, different collision energies,
different instrumental setups with different photon momentum coverage
and acceptances. So it is very difficult to say whether they are
consistent among themselves or not, whether they witness about the
same phenomenon, whether their contradictory claims are really
significantly inconsistent.
In this complicated situation we see only one way of pursuing
the matter: to use some theoretical model as a tool for connecting
various experiments. The guiding idea is that if a model is able,
after modulated by the experimental acceptances, to describe
several pieces of data without tuning its parameters for each
particular case, then the data are mutually consistent. It may
also indicate that the model is physically sound, but this need not
always be the case. The model may simply simulate an important
conventional contribution which was not properly taken into account
in every experimental analysis considered.

The purpose of the present paper is to assess the consistency of
various data on soft photon production in hadronic interactions
by using the glob model \cite{glob} and the modified soft
annihilation model \cite{MSAM} as interpolation tools.

In the next section, we deal with the glob model of Van Hove and the
present author. We first recapitulate its assumptions and equations,
as well as the way its parameters were chosen. Then we develop a
simpler, but physically equivalent mutation of the glob model, which
will be used throughout this paper. It can easily be implemented by
anyone wishing to perform her or his own calculations. Section 
\ref{MSAM}
deals similarly with the modified soft annihilation model of Thompson
and the present author. Unfortunately, here we are unable to offer a
user-friendly version. The central part of the paper is
Sec.~\ref{comparison}, where we show the results of model calculations
and their comparison with experimental data. 
Conclusions are summarized and commented upon in Sec.~\ref{comments},
where we also discuss the possible relation of anomalous soft photon
production to other phenomena.

\section{GLOB MODEL}
\label{glob}
\subsection{The original version}
\label{original}
As  mentioned in Sec.~\ref{intro}, we assume
that in some percentage of high-energy inelastic collisions,
a long-living, large, and dense system consisting of light
quarks, antiquarks, and gluons is formed. For the physical parameters 
of such a system, called glob, we will use the original estimates
\cite{vanhove,glob}: the glob mass $M_G=1$ GeV/$c^2$; the number of
partons within a glob $N_p=40$; the gluon/parton number ratio = 0.5;
and the light quark number ratio $u/d=1$.

Because of the fixed invariant energy, the momentum distribution of
partons is governed by the microcanonical distribution. The mean
number $n^G_\gamma$ of photons emitted by a glob is therefore given
by the expression
\begin{eqnarray}
\label{ngamma}
n^G_\gamma &=& k(M_G^2)\frac{t_0}{V_0}\sum_{i=2}^{N_p}\sum_{j<i}\int
\left[\left(p_i\cdot p_j\right)^2-m_i^2m_j^2\right]^{1/2} \nonumber\\
&\times & \frac{\sigma_{ij}(s_{ij})}{E_iE_j}\delta\left(\sum_{k=1}^
{N_p} E_k-M\right)\delta\left(\sum_{k=1}^{N_p}{\bf p}_k\right)
\prod_{k=1}^{N_p}\frac{d^3p_k}{E_k},
\end{eqnarray}
where ${\bf p}_k$ is the momentum of the $k$th parton in the glob rest
frame and $\sigma_{ij}(s_{ij})$ is the cross section for photon
production in head-on collisions of the $i$th and $j$th parton. The
processes $q + \overline{q}\rightarrow \gamma + g$ and
$g + q(\overline q) \rightarrow \gamma + q (\overline q)$ are
considered in the lowest order of QED and QCD.
The phase-space normalization constant is
\begin{equation}
\label{k}
\left[k(M_G^2)\right]^{-1}=\int
\delta\left(\sum_{k=1}^{N_p}E_k-M\right)
\delta\left(\sum_{k=1}^{N_p}{\bf p}_k\right)
\prod_{k=1}^{N_p}\frac{d^3p_k}{E_k}.
\end{equation}
The total cross section $\sigma_{ij}$ in
Eq.~(\ref{ngamma}) can be written as
\begin{equation}
\label{sigmaij}
\sigma_{ij}(s_{ij}) = \int dt_{ij}\frac{d\varphi_{ij}}{2\pi}
\frac{d\sigma_{ij}}{dt_{ij}},
\end{equation}
where $t_{ij}$ is the four-momentum transfer squared from one parton
to the photon in the collision of the $i$th and $j$th partons and
$\varphi_{ij}$ is the azimuthal angle of the photon in their rest 
frame.
After inserting (\ref{sigmaij}) into (\ref{ngamma}) and mapping the
independent variables onto a $(3N_p-2)$-dimensional unit cube
\cite{jadach}, we arrive at the equation that served in \cite{glob}
as the master equation for a Monte Carlo generator of the photon
momenta in the glob rest frame.

To transform the photon momentum from the glob rest frame to the
collision center-of-mass frame, we need to know the
glob momentum distribution in the latter. In \cite{glob} it was
assumed that the distribution in the glob rapidity and transverse
momentum squared
\begin{equation}
\label{globypt}
\frac{d^2N_G}{dy_Gdp^2_{T,G}}=N_GF_G\left(y_G,p_{T,G}^2\right)\ ,
\end{equation}
with the function $F_G$ normalized to unity, factorizes.
To proceed further, a gaussian glob rapidity distribution with
the width proportional to the maximum cms rapidity (i.e. $\log s$)
was chosen, with $\langle y_G^2\rangle^{1/2}=0.6$ for
$\sqrt s=11.51$ GeV (the BEBC experiment \cite{BEBC}).
The exponential $p^2_{T,G}$ distribution was assumed to be
collision energy independent with
$\langle p_{T,G}\rangle=0.3$~GeV/$c$.
In each Monte Carlo ``event", the rapidity and transverse momentum
were generated, increasing the number of independent variables
to $3N_p$. Then the photon momentum was transformed into the
collision center-of-mass frame.

To complete the model, the mean number of globs per event $N_G$ is
multiplied by the mean number of photons per glob $n_\gamma^G$
to give the mean number of photons per event $n_\gamma$.
Keeping in mind that
the cross sections of photon production subprocesses are proportional
to the strong coupling constant $\alpha_s$, the overall multiplication
constant that fixes the absolute normalization of photon yield is
\begin{equation}
\label{b}
B=\alpha_sN_G\frac{t_0}{V_0}\ .
\end{equation}
In order to fit the BEBC data \cite{BEBC}, its value was fixed in
\cite{glob} at $B=5\times10^{-3}\ c^{-1}$fm$^{-2}$. To get the
predictions for photon spectra at different collision energies or
with different incident particles it was suggested in \cite{glob}
to scale the mean number of globs $N_G$ according to empirically
known hadron multiplicity.

The first numerical realization of the glob model \cite{glob}, which
we have just briefly sketched, had serious disadvantages.
With $N_p$ fixed at 40, the dimension of the integral which was to
be evaluated by the Monte Carlo method was 120.
Determination of a detailed distribution in photon momentum with
low statistical errors thus became a computer-time-consuming task.
Also the computer code was complicated and difficult to use.
It was practically impossible to use the glob model as an event
generator for Monte Carlo studies of experimental setups.
In the next subsection we remove these drawbacks of the
original glob model \cite{glob} and suggest a computing scheme which
is easy to reproduce.

\subsection{A simplified version of the glob model}
\label{simplified}

In the glob model it is assumed that the internal properties of the
glob (total invariant energy of partons $M_Gc^2$, their number $N_p$,
as well as their momentum and flavor distributions) depend neither on
the type of process under study nor on the incident energy. This
allows us to construct a new version of the glob model, which is
physically equivalent to the original one, but much easier to use.

\begin{figure}
\begin{center}
\leavevmode
\setlength \epsfxsize{9cm}
\epsffile{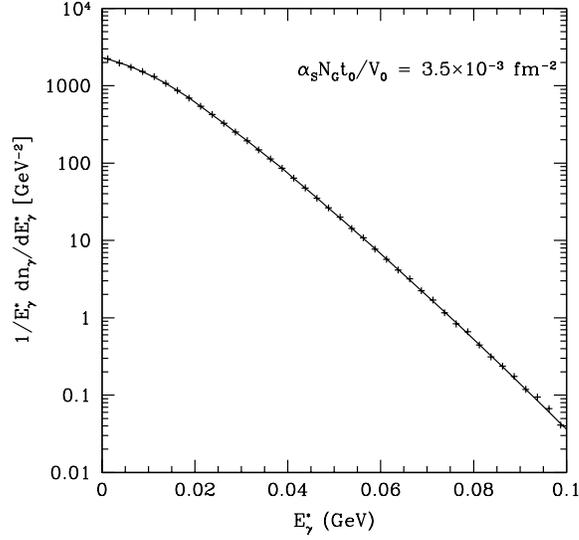}
\end{center}
\caption{Parametrization of the photon energy spectrum in the glob
rest frame (curve) and its comparison to the results of original Monte
Carlo calculation (crosses). }
\label{figure1}
\end{figure}

The invariant photon distribution in a general reference frame can
be written as a convolution of the photon
distribution in the glob rest frame with the momentum distribution
of globs in the general frame
\begin{equation}
\label{simplgen}
E_\gamma\frac{d^3n_\gamma}{d^3p_\gamma}=\int d^3p_G\frac{d^3N_G}
{d^3p_G}\ E^*_\gamma\frac{d^3n_\gamma^G}{d^3p_\gamma^*}\ .
\end{equation}
The asterisk refers to quantities in the glob rest frame and
\begin{equation}
{\bf p}^*_\gamma={\bf p}_\gamma+\frac{{\bf p}_G}{M_G}\left(
\frac{1}{E_G+M_G}{\bf p}_G\cdot{\bf p}_\gamma-E_\gamma\right)\ .
\end{equation}
According to the basic assumptions of the glob model, the photon
momenta are distributed isotropically in the glob rest frame. It is
thus sufficient to consider only the energy spectrum in the latter
frame. We can write
\begin{equation}
\label{egdist}
\frac{1}{E^*_\gamma}\frac{dn_\gamma^G}{dE^*_\gamma}=
\alpha_s\frac{t_0}{V_0}f(E_\gamma^*)\ .
\end{equation}
A study of this quantity in the framework of the original Monte
Carlo version of the glob model has showed that the function
$f(E_\gamma^*)$ can be parametrized as

\begin{equation}
\label{f}
f(E_\gamma^*)=a\exp\left\{-bE_\gamma^*-c(E_\gamma^*)^2\right\}\ ,
\end{equation}
with parameters $a$, $b$, and $c$ given in Table~\ref{abc}. Comparison
of the photon energy distribution in the glob rest frame based
on the parametrization (\ref{f}) with the original Monte Carlo
calculation in Fig.~\ref{figure1} shows that they are identical.

Introducing the distribution in the glob rapidity and transverse
momentum squared (\ref{globypt}) and using Eq.~(\ref{egdist}), we can
cast Eq.~(\ref{simplgen}) in the form
\begin{equation}
\label{d2ndpldpt}
\frac{d^2n_\gamma}{dp_{L,\gamma}dp_{T,\gamma}}=
\frac{B}{4\pi}\frac{p_{T,\gamma}}{E_\gamma}
\int F_G\left(y_G,p_{T,G}^2\right)f\left(E_\gamma^*\right)dy_G
\,dp_{T,G}^2 d\phi\ ,
\end{equation}
where $B$ is again given by Eq. (\ref{b}) and
\begin{equation}
\label{eg}
E_\gamma^*=\frac{1}{M_G}\left(E_GE_\gamma-p_{L,G}p_{L,\gamma}-
p_{T,G}p_{T,\gamma}\cos\phi\right)\ .
\end{equation}
Even if the integrals in Eq. (\ref{d2ndpldpt}) can be evaluated by
more conventional numerical methods, a Monte Carlo approach is
convenient to utilize, especially if the experimental acceptance
has to be taken into account.

\section{MODIFIED SOFT ANNIHILATION MODEL}
\label{MSAM}
The original soft annihilation model (SAM) \cite{SAM} was inspired by
Bjorken and Weisberg \cite{bjorweis} who suggested an
explanation of an unexpectedly large production of low-mass
lepton pairs discovered in many experiments \cite{llexps}.
In SAM, dileptons arise from an intermediate parton system (IPS)
by annihilation of quarks and antiquarks ($q\overline{q}\rightarrow
l^+l^-$ and $q\overline{q} \rightarrow l^+l^-gluon$) produced in
the initial stages of the reaction.
Unlike the IPS system of Van Hove \cite{vanhove}, the IPS of SAM
carries all the collision energy and is the only source of final
state hadrons. Its parameters are therefore fixed by data on the
production of hadrons \cite{lephad}. In 1981, SAM was compared
\cite{SAM81} to all the dilepton data available at that time and
appeared to be in satisfactory agreement with them.

Later on, the data on electron production at the CERN Intersecting
Storage Rings (ISR) were extended to lower values of the transverse
momentum \cite{AFSe}. The SAM was not able to follow a steep rise
of the $e^+/\pi$ ratio with decreasing transverse momentum.
The comparison of SAM (with the subprocess $q\overline{q}
\rightarrow \gamma + g$) to the EMC photon data was not without
flaws either. Both longitudinal and transverse spectra (in the
hadronic system rest frame) were flatter than the experimental one
\cite{EMCSAM}. To bring the calculations closer to reality, a gluon
component was introduced into the IPS in the modified soft
annihilation model \cite{MSAM} together with corresponding
subprocesses $g+q(\overline q)\rightarrow\gamma+q(\overline q)$ and
$g+q(\overline q)\rightarrow l^+l^-+q(\overline q)$.
With the quark and antiquark multiplicities
fixed at the same values as in SAM (in order not to change outgoing
hadron multiplicities), the mean energies of partons became smaller,
which also made the spectra of dileptons and photons softer.
This improved the agreement not only with the results of \cite{EMC}
and \cite{AFSe}, but also with some older dilepton data (see
\cite{MSAM}). We refer the reader to \cite{MSAM} for technical details
of the model, which we will use also in this work. Let us only note
that the model, when applied to soft photon production, does not
contain any free parameters. For each projectile, target, and incident
energy combination, the mean number of partons in IPS was fixed by the
mean hadron multiplicity.

\section{Comparison of models to data}
\label{comparison}

In this section we compare the outcomes of the glob model \cite{glob}
and the modified soft annihilation model (MSAM) \cite{MSAM} with
existing data. For data which provide absolute normalization, the
procedure is straightforward. They are those listed in the 
Introduction except for the AFS \cite{AFS} and Anto\v{s} {\it et al.} 
\cite{antos} experiments.

In paper \cite{AFS}, the AFS Collaboration presented the
$p_T$-spectrum of photons observed in $pp$ collisions at $\sqrt s=
63$~GeV (their Fig.~13) normalized to expectations. The latter were,
in turn, given in terms of ratios to the $(\pi^++\pi^-)/2$ yields (AFS
Fig. 12). The comparison of models to their results would be possible,
but would require the recall of information from other experiments and
the inclusion of the AFS efficiences and acceptance. As we have
already mentioned, the AFS Collaboration  found their data compatible
with the other data known at the time \cite{SLAC,BEBC}, which will be
explored here in detail. We would be unlikely to add much to their
statement even if we compared \cite{AFS} with our models.

With the data of Anto\v{s} {\it et al.}, the situation is different.
At first glance they seem to be incompatible with all the experiments
that have reported the existence of the soft photon anomaly. It is
therefore very important to include results by Anto\v{s} {\it et al.}
in our study. But in the paper \cite{antos} they presented the results
only in ``arbitrary units". A plausible way of comparing models to
data \cite{antos} was suggested to us by J.~Schukraft. It is described
in the relevant subsection below.

Unlike in the original paper on the glob model \cite{glob}, more
detailed and statistically more precise results of the simplified
version of the glob model can now be combined with the other
independent sources of photons (MSAM, bremsstrahlung calculated by
experimentalists on the basis of empirical charged hadron
distributions or hadron production models) producing a total
theoretical expectation. The latter is then compared to the
experimental data. We can thus also determine the basic parameter of
the glob model $B=\alpha_SN_Gt_0/V_0$ more reliably.

\subsection{BEBC experiment by Chliapnikov \protect{\it et al.}
\protect\cite{BEBC}}
\label{BEBCcomp}

\begin{figure}
\begin{center}
\leavevmode
\setlength \epsfxsize{9cm}
\epsffile{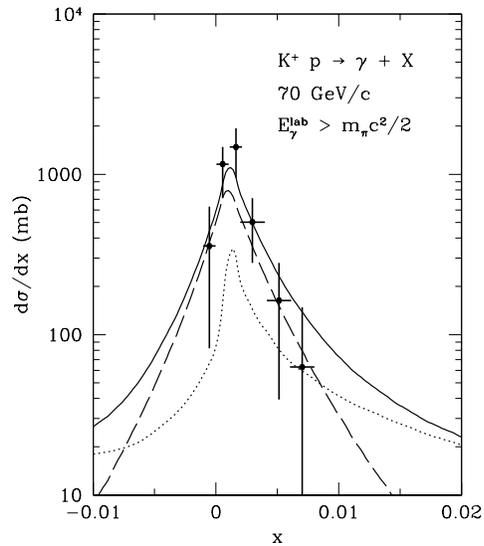}
\end{center}
\caption{BEBC data \protect\cite{BEBC} compared to the classical
bremsstrahlung formula (dotted), the glob model (dashed), and their
sum (solid). }
\label{figure2}
\end{figure}

We start with the BEBC data \cite{BEBC} because they were used in
\cite{glob} to fix the absolute normalization of the glob model.
The experimental cut $E_\gamma^{lab}>m_\pi c^2/2$ was enforced
in computations within both models, as in bremsstrahlung
calculations \cite{BEBC}. In Fig.~\ref{figure2} we show the  data on
the distribution of photons in $x=2p_{L,\gamma}/\sqrt s$ together
with the leading term bremsstrahlung calculation from \cite{BEBC}
and with the outcome of the glob model. The total theoretical
expectation, which should be compared to data, is the sum of the
glob and bremsstrahlung curves. In the glob model calculation we
used the value $B=3.5\times10^{-3}\ c^{-1}$fm$^{-2}$ of the overall
multiplication constant (\ref{b}). The present value of $B$, which
is 0.7 times the value used in \cite{glob}, was
chosen to give good agreement of the theoretical $x$-spectra not only
with the BEBC results, but also with the data \cite{NA22} on prompt
photon production in $K^p$ and $\pi^+p$ interactions at 250 GeV/$c$
(see below).

The reader has certainly noticed that the MSAM curve is not shown
in Fig.~\ref{figure2}. Neither are the MSAM photons included into the
total theoretical expectation, which should be compared with data.
The reason is that their $x$-distribution is almost flat on the
scale of Fig.~\ref{figure2} and over a wider range it resembles the
spectrum of photons from $\pi^0$ decays (not shown). It was the virtue
of the experimental procedure used in \cite{BEBC} for isolating the
anomalous component in the $x$-spectrum that the photons with a
spectrum similar to that from the $\pi^0$ decays were subtracted from
the total yield. In this way, the MSAM photons were also subtracted
and are not present in the data points shown in Fig.~\ref{figure2}.

The transverse momentum spectra of photons produced in 70 GeV/$c$
$K^+p$ collisions are depicted in Fig.~\ref{figure3}. We can see that
the data above $p_T\approx 15$ MeV/$c$ are well described by the
superposition of both models and bremsstrahlung. The glob model
provides the most important contribution up to $p_T\approx
45$ MeV/$c$, where MSAM takes over. The classical bremsstrahlung
formula is dominant below $p_T\approx5$ MeV/$c$ (as expected from the
Low formula), but even it is unable to account for all the
experimentally observed yield. For $p_T<10$ MeV/$c$, the mean excess
of data above the total expectation (full curve) is roughly equal to
the total expectation itself.

\subsection{SLAC experiment by Goshaw \protect{\it et al.}
\protect\cite{SLAC}}

The inclusive $p_T^2$-distribution of photons from $\pi^+p$
interactions at 10.5 GeV/$c$ is shown in Fig.~\ref{figure4}. The
central values of all but one data point lie above the dotted curve,
which represents a classical bremsstrahlung calculation. But,
accounting for experimental errors, this does not represent a
statistically significant discrepancy.

To perform a glob model calculation, we have to rescale the
multiplicative constant $B$ to the new energy. In \cite{glob} it was
suggested that it should be done according to the mean hadron
multiplicity. But we feel that at this very low energy ($\sqrt s
=4.54$~GeV) this is inadequate. Due to the energy-momentum constraints
it is much more difficult to produce a glob with mass 1~GeV in
addition to a baryon in the final state than a pion (pions account for
most of the produced multiplicity) with a mass roughly seven times
smaller. The dependence of $N_G$ on the collision energy should be, at
small energies, much steeper than that of overall hadron multiplicity
and should behave more like, let us say, the mean multiplicity of
centrally produced $\phi$(1020)'s. Guided by the experimental results
on the production of the latter in $pp$ collisions (see, e.g., Fig.~3
in \cite{otwinowska}), we assume that the glob multiplicity at
10.5~GeV/$c$ is three times smaller than that at 70 GeV/$c$. In this
way we mimic the threshold effect which must exist in the production
of globs in low-energy collisions. Scaling according to the ratio of
produced charged hadron multiplicities in 70~GeV/$c$ $K^+p$ and
10.5~GeV/$c$ $\pi^+p$ collisions would lead to a decrease by a little
smaller factor of 2.1. The ratio of charged multiplicities is 1.6.

In order to be able to compare the model predictions with the data
we have taken into account the experimental cuts $0<x<0.01$
and $E_\gamma^{lab}>30$ MeV, number of events (33,676), and
detection efficiency (0.25) given in \cite{SLAC}. The results of the
glob model are depicted by a dashed curve in Fig.~\ref{figure4}, the
results of MSAM lie below the lower edge of the diagram. The sum of
all three theoretical components (solid curve) is now higher than the
central values of most of the data points, but is as compatible with
them as was the pure bremsstrahlung component. We can thus conclude
that even if the results of the SLAC experiment \cite{SLAC} do not
require any additional mechanism rather than classical bremsstrahlung,
they do not exclude an additional contribution of the size given by
the glob model. The glob mechanism dominates over bremsstrahlung for
$p_T \gtrsim 15$ MeV/$c$.

\subsection{EMC experiment by Aubert \protect{\it et al.}
\protect\cite{EMC}}

This experiment, which measured the inclusive photon production in
deep inelastic scattering of 200 GeV/$c$ muons in hydrogen, is
usually included into a common list with experiments on anomalous
soft photon production in hadronic collisions. We will show here
that the kinematic range and probably also the production
mechanism of the anomaly observed in \cite{EMC} are different from
those reported by hadronic experiments \cite{BEBC,NA22,SOPHIE}.

The inclusive photon distributions in \cite{EMC} are normalized to
all deep inelastic events. We will deal with two of them that are
presented in transverse momentum $p_T$ to the virtual photon current
and fractional energy $z$, defined as $E_\gamma^{lab}/\nu$, where
$\nu$ is the energy lost by the scattered muon. Because we did not
have access to empirical distributions in $\nu$ and in the
invariant energy of hadronic system $W$, we fixed them at their
mean values $\langle \nu\rangle=113$~GeV,
$\langle W^2\rangle=195$~GeV$^2$, quoted in \cite{EMC}. When
calculating the transverse momentum distribution, we applied
the cut $z>0.05$. In virtue of both models we are using here,
we assume that the intermediate parton systems they are dealing
with are bound to the rest frame of the hadronic system produced
in deep inelastic scattering.

The $z$-distribution of photons from the glob model and MSAM (taken
from \cite{MSAM}) is compared with the data in Fig.~\ref{figure5}. We
can see that the photons from the glob model are completely irrelevant
here, because their energies are much smaller than those of the prompt
anomalous photons observed in \cite{EMC}. Having so little energy, the
glob photons are below the $z$-cut and do not appear at all in
Fig.~\ref{figure6}, which shows the $p_T^2$ distribution of anomalous
prompt photons. The glob model alone provides a successful description
of data in both cases.

This points out to a completely different production mechanism of the
EMC anomalous photons. Rather than being generically related to the
anomalous soft photons in hadronic collisions, they have more in
common with the trimuons discovered a long time ago \cite{trimu}. It
is indicated by the fact that also the trimuon production was
satisfactorily described by the soft annihilation model
\cite{trimusam}.

\subsection{EHS-NA22 experiment by Botterweck \protect{\it et al.}
\protect\cite{NA22}}
\label{NA22comp}

EHS-NA22 Collaboration studied inclusive cross sections of prompt soft
photon production in $K^+p$ and $\pi^+p$ interactions at 250 GeV/$c$.
In our calculations within the glob model and MSAM we applied the
cut $E_\gamma^{lab}>m_\pi c^2/2$, as introduced in the experiment.

The models are compared with the experimental inclusive photon
production cross section in $x=2p_L/\sqrt{s}$  in Figs.~\ref{figure7}
($K^+p$ collisions) and \ref{figure8} ($\pi^+p$ collisions). The
contribution from MSAM is not included for the same reasons as in
subsection \ref{BEBCcomp}. The agreement between data and the combined
theoretical expectation is very good. Of course, keeping in mind the
inevitably large experimental errors (each data point was obtained as
a difference of two big numbers--the total yield minus the calculated
yield from the radiative decays of hadrons), theoretical curves scaled
down by a factor of, say, $\approx 1.5$ would be also acceptable.

In Figs.~\ref{figure9} and \ref{figure10}, the measured inclusive
differential cross sections in transverse momentum are presented
together with the classical bremsstrahlung estimate made by
experimenters, two model curves, and a sum of these three components.
For both $K^+p$ and $\pi^+p$ initial states, the theoretical
expectation does not match the data very well.
It is below the data for ultrasoft photons ($p_T\lesssim 10$ MeV/$c$)
and overshoots the data in the medium region (10~MeV/$c \lesssim
p_T\lesssim 50$~MeV/$c$). The former feature is common with other
hadronic experiments and will be discussed later. The latter may have
several origins. First of all, the procedure of extrapolations to
different collision energies suggested in \cite{glob} may be
unreliable. This interpretation is somewhat called into question by
the good agreement with the results of the WA83/SOPHIE experiment
\cite{SOPHIE} (see below), which was done at even a slightly
higher collision energy ($p_{lab}=280$~GeV/$c$). Another reason may
lie in our inadequate simulation of experimental conditions. Besides
the energy cut mentioned above, the experimenters introduced several
others in an effort to minimize systematic errors. The latter would be
very difficult to implement in our model calculations, because such a
task would require a detailed knowledge about the experimental setup
and a model for the conversion of photons to electrons in the metal
foils placed inside the Rapid Cycling Bubble Chamber. We are not, of
course, able to assess here how reliably it was possible to correct
for all those cuts, and to what extent they persist in the final
experimental distributions.

There is another important issue to be discussed in connection with
the paper \cite{NA22}. In their Figs. 7a,b and 8a,b, the authors of
\cite{NA22} show the predictions of the glob model, which differ a
little from what we presented here. In fact, their model histograms
are lower than our curves and provide a better description of
differential cross sections in $p_T$. Unfortunately, the model
predictions in \cite{NA22} were obtained under oversimplified
assumptions. The model histograms \cite{glob} for $K^+p$ collisions at
70~GeV/$c$ were only multiplied by the ratios of total inelastic cross
sections at 250 and 70 GeV/$c$. The authors of \cite{NA22} thus
neglected the change of mean glob multiplicity with the collision
energy and assumed that the form of distributions is collision-energy
independent, $x$ and $p_T$ were treated as ``scaling variables".

\subsection{SOPHIE/WA83 experiment by Banerjee \protect{\it et al.}
\protect\cite{SOPHIE}}
\label{SOPHIEcomp}

The high statistics study \cite{SOPHIE} of direct soft photon
production in $\pi^-p$ collisions at 280 GeV/$c$ was based on a sample
of 310,390 events observed in the apparatus consisting of the OMEGA
spectrometer and two electromagnetic calorimeters. The results are
given in the numbers of photons per bin in the variable under
consideration ($p_T$ or $E_\gamma^{lab}$). The numbers have been
corrected for the $\gamma$ detection efficiency, but no attempt has
been made to correct for the geometrical acceptance and extract the
differential cross sections. Generally speaking, for experiments with
a non-trivial geometrical acceptance that covers only a part of the
phase space, such a way of presenting results is least biased and
therefore most valuable \cite{DLS}. Any attempt to go beyond it would
require some assumptions about either the production mechanism or the
photon distribution in inaccessible phase-space regions. In order to
compare a model with data, one has to modulate theoretical
distributions with experimental acceptance.

We took full advantage of the possibilities provided by the simplified
version of the glob model and merged it with a program that described
the geometrical acceptance of the WA83 experiment\footnote{
I am indebted to M. Spyropoulou-Stassinaki for providing me with the
necessary information and to A. Belogianni for checking the relevant
part of my computer code.} in order to calculate the same sort of
distributions as shown in \cite{SOPHIE}. Such a project
would be very difficult to accomplish with a very inefficient photon
generator based on the original version. For MSAM, the acceptance
modulated calculations do not represent a novelty \cite{SAM81,MSAM}.

The photon transverse momentum spectra from models are compared to the
experimental one in Fig. \ref{figure11}. Again, a reasonable agreement
has been achieved for $p_T\gtrsim 10$ MeV/$c$. The transition from the
glob model to MSAM is now located at $p_T\approx 30$ MeV/$c$. It is
interesting that the size of the classical bremsstrahlung is roughly
equal to that of the model which just dominates. The excess of data
over the total theoretical expectation in the ultrasoft region
($p_T\lesssim 10$~MeV/$c$) is huge, about sevenfold for the lowest
data point.

\subsection{Soft photon experiment by Anto\v{s} \protect{\it et al.}
\protect\cite{antos}}
\label{antoscomp}

As we have already mentioned, the experiment measured the photon
production in $p$Be interactions at 450 GeV/$c$. For those who want
to compare its results with theoretical estimates, the missing
absolute normalization is an obstacle. J.~Schukraft, member of both
the HELIOS Collaboration \cite{HELIOS} and the experimental group
\cite{antos}, suggested to us to fix the model normalization by
comparing the bremsstrahlung estimates in \cite{HELIOS} and
\cite{antos}. Because of similarities between the two apparatus, they
should be identical. But the former is given in terms of the double
differential cross section, the latter in ``arbitrary units". This
allowed us to recalculate the cross sections provided by models to the
``arbitrary units" of the experiment \cite{antos}. For this purpose,
we used the Be results of Fig.~13 from the Schukraft
presentation at the Quark Matter '88 conference \cite{HELIOS}.

Figure \ref{figure12} shows the $p_T$ distribution of photons produced
with zero rapidity in the cms of proton-nucleon center-of-mass system
after subtraction of the decay background. The data taken by both
detection methods are shown. The systematic errors of the data,
the decay background and the bremsstrahlung calculation are not
reproduced from the original Fig. 5a to keep our figure uncluttered.
The total theoretical expectation agrees nicely with the results from
the BaF$_2$ detector and is compatible, taking into account the large
systematic errors, with the BGO array results. In the kinematic region
of the experiment \cite{antos}, the MSAM
contributes very little. The glob model provides the most important
contibution in the medium region 8~MeV/$c \lesssim p_T \lesssim
40$~MeV/$c$. There is practically no excess in data over the
theoretical expectation in the ultrasoft region.

\section{Conclusions and comments}
\label{comments}

We have seen in the previous section that in the prompt photon
production it is useful to distinguish among three different
kinematical regions:

(1) ultrasoft, with photon transverse momenta less than, say,
10 MeV/$c$;

(2) very soft, 10 MeV/$c \lesssim p_T \lesssim 50$~MeV/$c$; and

(3) soft, characterized by $p_T \gtrsim 50$~MeV/$c$.\\
Of the various components that we included in our phenomenological
approach, the classical bremsstrahlung formula is invincible in the
ultrasoft region. In the very soft region, the main contribution to
the total theoretical expectation comes from the glob model
\cite{glob}. The yield from the modified soft annihilation model
\cite{MSAM} peaks somewhere beyond the range explored by the hadronic
experiments considered here and is therefore the only theoretical
component which rises with $p_T$. Whether and where it will be 
above the remaining two components (classical bremsstrahlung and glob
model) depends on the incident energy and the setup of the experiment. 
But it usually dominates in the soft region.

The experimental data is mixed, with some experiments showing a much
larger excess over the hadronic decay background than expected from
their classical bremsstrahlung calculation, and other experiments
claiming agreement between the excess and bremsstrahlung. The main
result of this work is that, for all the experiments  the observed
$x$- and $p_T$- distributions of direct photons are reasonably well
described above $\approx 10$~MeV/$c$ (that is except for the ultrasoft
transverse momentum region), by a mixture consisting of the classical
bremsstrahlung calculation (taken from original experimental papers)
and two theoretical models. The latter are detailed enough to include
experimental cuts and acceptances. The rules for extrapolating them to
different energies have clearly been stated beforehand. On the basis
of agreement of all data with the theoretical expectation, we conclude
that all the experimental data (except for the $p_T$-distributions in
the ultrasoft region) on anomalous soft photon production are mutually
consistent.

\begin{figure}
\begin{center}
\leavevmode
\setlength \epsfxsize{9cm}
\epsffile{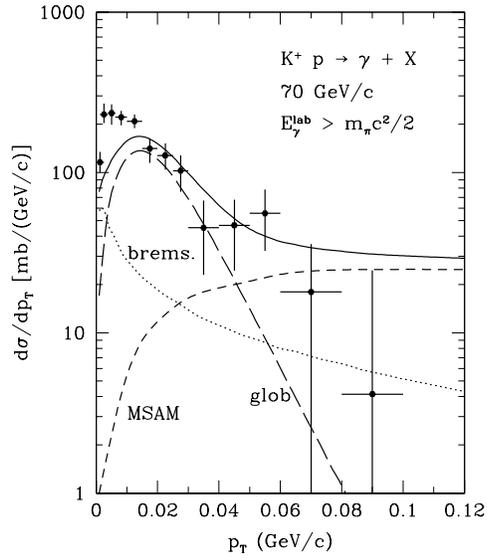}
\end{center}
\caption{BEBC data \protect\cite{BEBCpt} and their comparison with the
sum (solid) of the classical bremsstrahlung formula, the glob model,
and the modified soft annihilation model (MSAM).  }
\label{figure3}
\end{figure}

\begin{figure}
\begin{center}
\leavevmode
\setlength \epsfxsize{9cm}
\epsffile{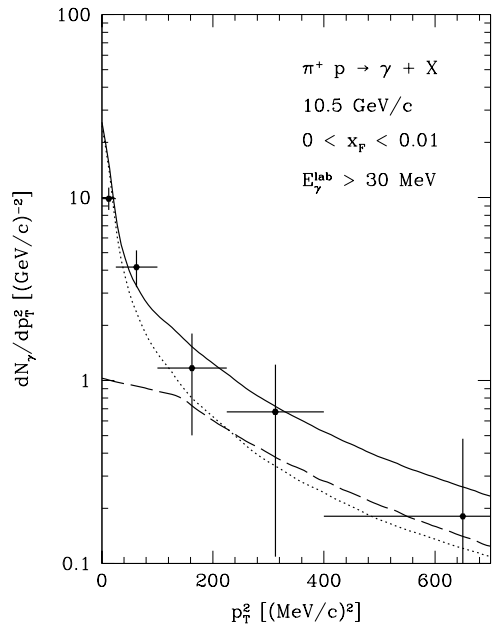}
\end{center}
\caption{Goshaw \protect{\it et al.} data \protect\cite{SLAC} compared
to the classical bremsstrahlung calculation (dotted),
the glob model (dashed), and their sum (solid).}
\label{figure4}
\end{figure}

\begin{figure}
\begin{center}
\leavevmode
\setlength \epsfxsize{9cm}
\epsffile{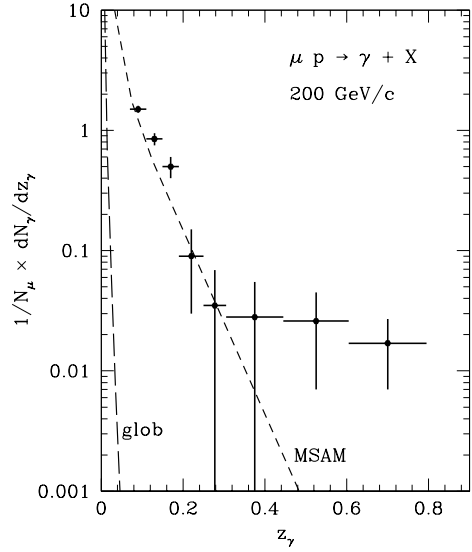}
\end{center}
\caption{EMC data \protect\cite{EMC} compared to the modified soft
annihilation model (MSAM) and the glob model. The rightmost data
points can be explained as a bremsstrahlung from the scattered muon
\protect\cite{EMC}. }
\label{figure5}
\end{figure}

\begin{figure}
\begin{center}
\leavevmode
\setlength \epsfxsize{9cm}
\epsffile{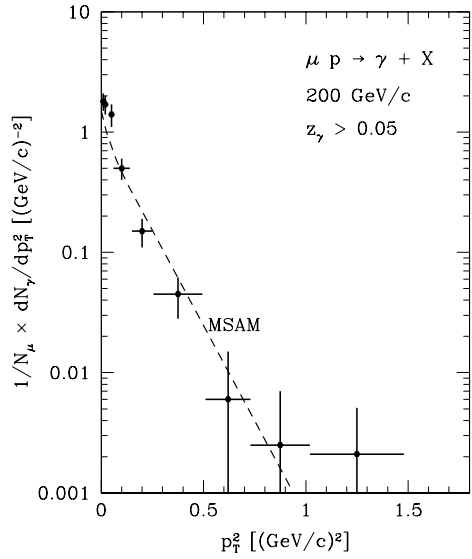}
\end{center}
\caption{EMC data \protect\cite{EMC} compared to the modified soft
annihilation model (MSAM). Because of the \protect$z$-cut, the
radiation from globs does not contribute at all (compare
Fig.~\protect\ref{figure5}). }
\label{figure6}
\end{figure}

\begin{figure}
\begin{center}
\leavevmode
\setlength \epsfxsize{9cm}
\epsffile{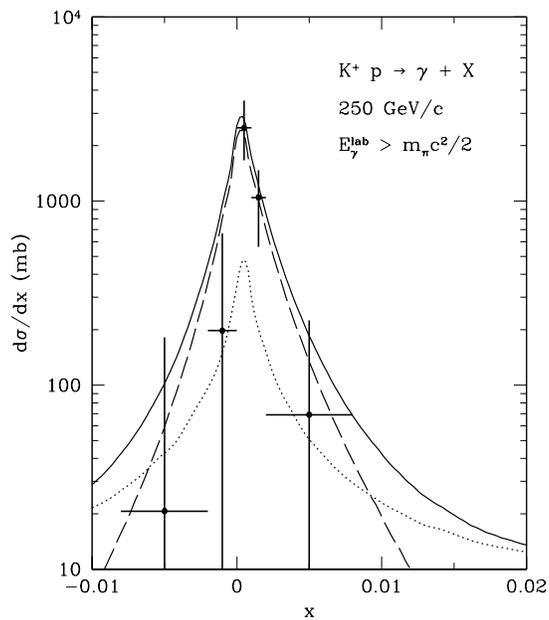}
\end{center}
\caption{NA22 \protect$K^+p$ data \protect\cite{NA22} compared
to the sum (solid) of
the classical bremsstrahlung calculation (dotted) and the glob
model (dashed). }
\label{figure7}
\end{figure}

\begin{figure}
\begin{center}
\leavevmode
\setlength \epsfxsize{9cm}
\epsffile{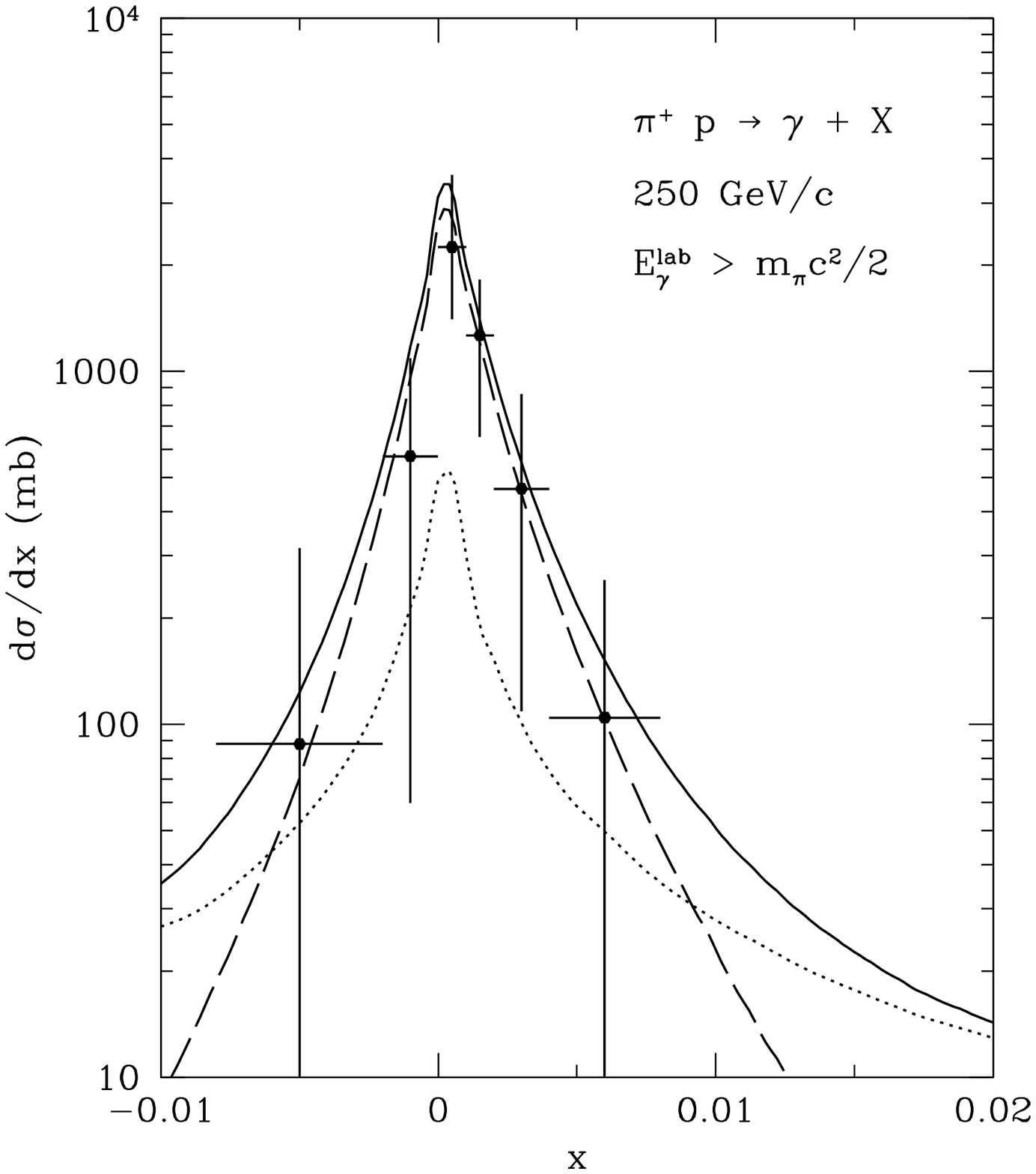}
\end{center}
\caption{Same as Fig. \protect\ref{figure7}, but with a \protect$\pi^+$
projectile.}
\label{figure8}
\end{figure}

It should be stressed that the magnitudes of different components in
the phenomenological mixture did not come out as a result of fitting
the experimental data, but are given as an interplay between their
physics properties and experimental conditions (incident energy,
instrumental cuts and acceptances). The actual numbers may be very
different in different cases. For example, in the $p_T$ distribution
from the BEBC experiment \cite{BEBCpt}, the maximum
glob/bremsstrahlung ratio is almost seven (see Fig.~\ref{figure3}),
whereas in the $p$Be collisions at 450~GeV/$c$ \cite{antos}, it does
not exceed four (Fig.~\ref{figure12}), and is able to squeeze into the
empirical upper limits of direct photons provided by the BaF$_2$
method in \cite{antos} (their Fig. 6). In the former experiment, the
MSAM gives the dominant contribution for $p_T\gtrsim 50$~MeV/$c$, but
is below the bremsstrahlung up to the highest $p_T$'s in
Fig.~\ref{figure12}. In the WA83/SOPHIE experiment \cite{SOPHIE}, the
size of bremsstrahlung in the very soft and soft regions is roughly
equal to that of the dominating model (glob or MSAM,
see Fig.~\ref{figure11}).

\begin{figure}
\begin{center}
\leavevmode
\setlength \epsfxsize{9cm}
\epsffile{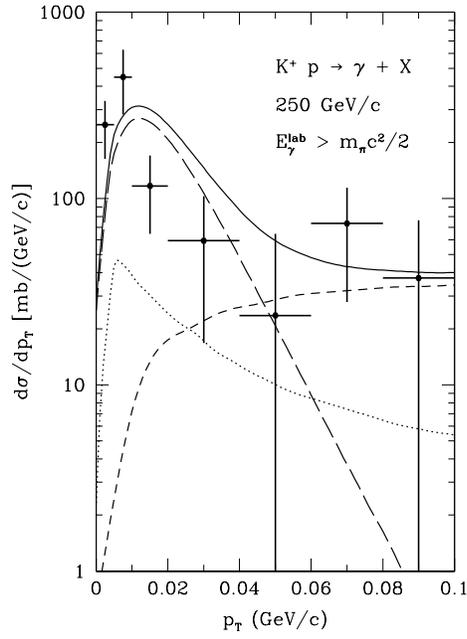}
\end{center}
\caption{NA22 \protect$K^+p$ data \protect\cite{NA22} and their
comparison with
the sum (solid) of the classical bremsstrahlung formula
(dotted), the glob model (long dash),
and the modified soft annihilation model (short dash).  }
\label{figure9}
\end{figure}

\begin{figure}
\begin{center}
\leavevmode
\setlength \epsfxsize{9cm}
\epsffile{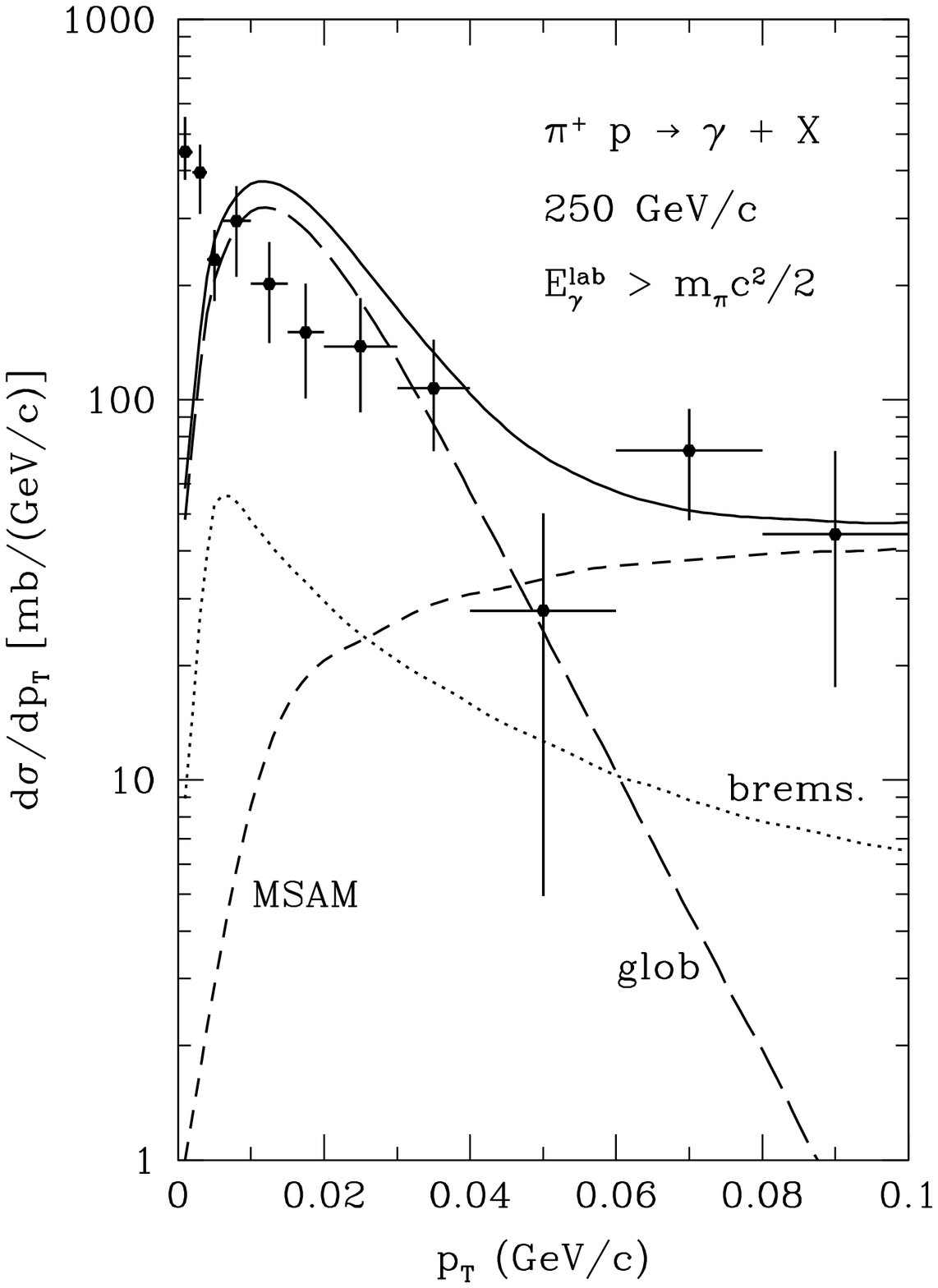}
\end{center}
\caption{Same as Fig. \protect\ref{figure9}, but with a \protect$\pi^+$
projectile.}
\label{figure10}
\end{figure}

\begin{figure}
\begin{center}
\leavevmode
\setlength \epsfxsize{9cm}
\epsffile{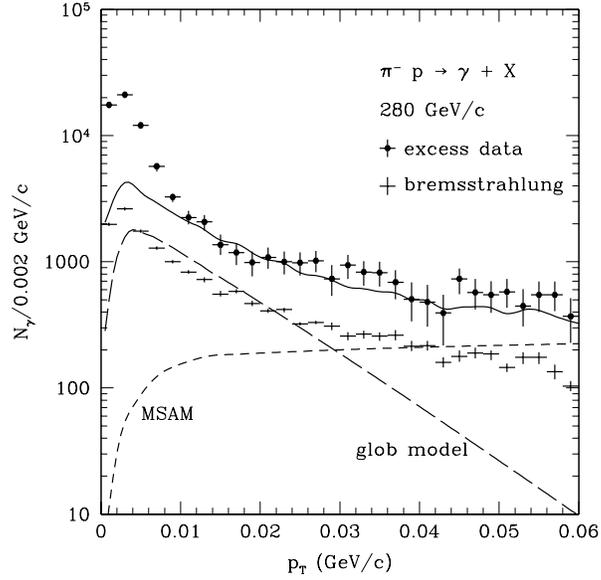}
\end{center}
\caption{Comparison of the excess in WA83 data \protect\cite{SOPHIE}
over known sources with the sum (solid curve) of the photon yields
from classical bremsstrahlung, glob model, and modified soft
annihilation model (MSAM).}
\label{figure11}
\end{figure}

\begin{figure}
\begin{center}
\leavevmode
\setlength \epsfxsize{9cm}
\epsffile{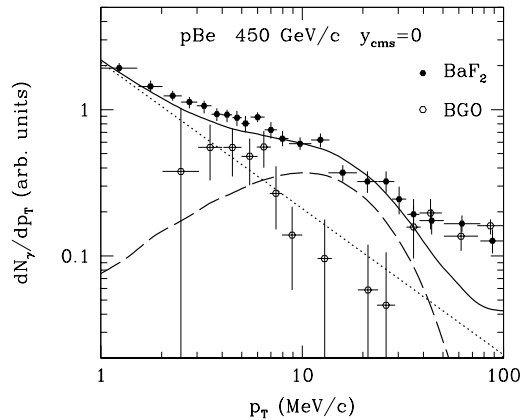}
\end{center}
\caption{450 GeV/\protect$c$ \protect$p$-Be direct photon data
\protect\cite{antos} compared to the sum (solid curve) of
the classical bremsstrahlung calculation (straight line), the glob
model (long dash), and MSAM (under the scale). }
\label{figure12}
\end{figure}

Let us turn now to the ultrasoft region, where some experiments agree
with the theoretical expectation (completely dominated and therefore
represented here exclusively by the classical bremsstrahlung formula),
whereas others see a significant excess. While the delicacy of the
experiments is underlined by the juxtaposition of the early HELIOS
\cite{HELIOS} results and those of \cite{antos}, and the differing
results could lie in differing experimental techniques, the apparent
contradiction between the experiments might also be due to different
underlying physics conditions. Without any theory or model able to
describe the anomalous excess, the field is open to speculations. Here
is one possibility:

The data seem to suggest (compare Figs.~\ref{figure3},
\ref{figure9}, \ref{figure10}, \ref{figure11}, and \ref{figure12}) that the
ultrasoft excess decreases with increasing mass of the projectile.
There is one experiment that seems to contradict this suggestion:
the historically first photon experiment \cite{SLAC}, which did not
see anything anomalous with pions. But let us recall that this
is the only experiment which used, besides the laboratory
energy cut $E_\gamma^{lab}>30$~GeV, also the cut on the longitudinal
photon momentum in the cms $0 < 2p_{L,\gamma}/\sqrt{s} < 0.01$. The
combination of those two cuts suppresses the yield of low-$p_T$
photons, especially if they are produced in a narrow cone around
the projectile momentum. So the absence of the excess in this case
need not mean its true nonexistence.

The regularity above may imply that the excess is caused by
bremsstrahlung from the projectile experiencing a (multiple)
small-angle scattering. If we ignore instrumental effects (residual
gas, thick target, stray photons from background interactions), we can
think, e.g., about (multiple) soft gluon exchange between projectile
and target before the hard, multiparticle production interaction takes
place.

Of course, there is still a possibility that the experimenters who saw
the excess significantly greater than the classical bremsstrahlung in
the ultrasoft region neglected some important conventional
contribution.

In the context of the ultrasoft region, it must be also noted that the
classical bremsstrahlung formula \cite{jackson}, which was used by all
experimenters to assess the expected level of photon production, is a
big unknown. It has been shown a long time ago that it should be valid
also in the quantum world in the limit of negligibly small photon
momenta. It enables one to estimate the photon yield if the cross
section of the corresponding nonradiative reaction is known. For
higher photon energies, the non-leading terms in the Low expansion
\cite{low} become important. But they cannot be evaluated without a
more detailed knowledge of the underlying strong dynamics of the
collisions. It is not clear what is the region of validity of the
classical approximation. In some examples, see, e.g., \cite{brems},
it is very narrow.

As already stated in Sec.~\ref{intro}, the agreement of the models
with experimental data need not imply that the mechanisms of photon
productions they are based on are real. But let us assume for a moment
that the models we used in this work\footnote{In fact, we do not know
about any other model which would have been compared in detail with a
broad set of data.} are more than a clever parametrization of all
existing data, that they explain the very origin of additional
photons. We can then go beyond mere phenomenology and address, at
least qualitatively, two important issues, which may have experimental
implications.

The first remark concerns the dependence of the prompt photon yield on
the associated hadron multiplicity. The prediction \cite{square}
of faster than linear dependence in dilepton production, and the
experiment that seemed to observe it \cite{afssquare} evoked a false
impression that this effect must take place wherever an anomalous
electromagnetic signal is encountered. As discussed in more
detail in \cite{plpitt} (and, to some extent, already in
\cite{square}), the actual behavior depends on the production
mechanism. We expect a roughly linear dependence if the dominant
mechanism is bremsstrahlung and faster than linear dependence for the
(modified) soft annihilation model \cite{square}. In the case of
photon production from the glob model, the intermediate parton system
represents only a part of the event and most of the final state
hadrons do not originate from it. We therefore expect no correlation
between the very soft photon production rate and the associated hadron
multiplicity.

Let us also note that a simultaneous observation of two (or more)
ultrasoft effects would be a nice confirmation of Van Hove's glob
mechanism \cite{vanhove}. It may manifest itself, for example, by
stronger short-range correlations among pions in events with very soft
photons.

The photon production in high energy collisions remains intriguing and 
lacking complete explanation and therefore deserves continuing 
experimental and theoretical attention.

\acknowledgements

I am indebted to P.~Braun-Munziger, C.~Fabjan, U.~Goerlach, 
J.~Kapusta, O.~Nachtmann, J.~Pi\v{s}\'{u}t, M.~Prakash, 
J.~Schukraft, E.~Shuryak, P.~Sonderegger, M.~Spyropoulou-Stassinaki, 
J.~Stachel, J.~Thompson, X.-N. Wang, and C.~Woody
for stimulating discussions. It must be noted, however, that not
all of them share the views expressed in this work.
Special thanks are due to J.~Thompson, who helped me to clarify my
understanding of the experimental situation, read the manuscript
very carefully, and suggested many improvements of the text.
My interest in soft photon production, like that of many others,
was stimulated by the late Yves Goldschmidt-Clermont.
The collaboration with the late L\'{e}on Van Hove was very pleasant
and profitable for me.
The stay at the State University of New York at Stony Brook
was supported by the U.S. Department of Energy under grant
No.~DE-FG02-88ER-40388.
This work was initiated when I participated in the program Strong
Interactions at Finite Temperatures at the Institute of Theoretical
Physics, University of California at Santa Barbara. My stay there 
was made possible by the National Science Foundation under the 
Grant No. PHY89-04035.  
A part of the work was done during a visit to the CERN Theory 
Division, the hospitality of which is gratefully acknowledged.

\begin{table}
\caption{Parameters of the photon energy distribution in the glob rest
frame}
\begin{tabular}{cccc}
$E_\gamma^G$ (GeV) & $a$ (fm$^{-2}$GeV$^{-2}$) & $b$ (GeV$^{-1}$) &
$c$ (GeV$^{-2}$) \\
\hline
 $< 0.02$     & $6.60\times10^5$ & $3.39\times10^1$ &$1.61\times10^3$
 \\
 (0.02, 0.04) & $1.05\times10^6$ & $8.15\times10^1$ &$4.04\times10^2$
 \\
 $> 0.04$     & $1.63\times10^6$ & $1.02\times10^2$ &$1.82\times10^2$
 \\
\end{tabular}
\label{abc}
\end{table}

\end{document}